# Effect of Magnetic Field on the Damping Behaviour of a Ferrofluid Based Damper

Durga N K P Rao Miriyala
Department of Mechanical Engineering
Pillai College of Engineering
New Panvel – 410206, India
mdurgarao@mes.ac.in

P S Goyal
Pillai College of Engineering
Department of Electronics
New Panvel – 410206, India
psgoyal@mes.ac.in

*Abstract*— **This paper is an extension of our earlier work where we had reported a proof of concept for a ferrofluid based damper. The damper used ferrofluid as damping medium and it was seen that damping efficiency of the damper changes on application of magnetic field. The present paper deals with a systematic study of the effect of magnetic field on the damping efficiency of the damper. Results of these studies are reported. It is seen that damping ratio varies linearly with magnetic field (ζ / H = 0.028 per kG) for magnetic field in range of 0.0 to 4.5 kG.
It may be mentioned that ferrofluid is different from magnetorheological fluid even though both of them are magnetic field-responsive fluids. The ferrofluid-dampers are better suited than MR Fluid-dampers for their use in automobiles.**

*Keywords—active dampers, ferrofluids, damping ratio, magnetic field*

## I. INTRODUCTION

Cars, buses and other vehicles, often, experience random oscillatory motions because of uneven roads or pot holes etc., and these oscillations are damped using shock absorbers or dampers. In general, above dampers work on piston-cylinder principle and their damping characteristics depend on the damping medium (e.g., oil) [1]. Contrary to what are available, one would like to have dampers whose damping behaviour can be controlled externally. This is because if the fact that damping requirements of a vehicle depend on road condition and load etc. The development of Magnetorheological (MR) Dampers has been a major achievement in field of shock absorbers [2, 3]. The damping behaviour of above dampers can be controlled by applying external magnetic field. These dampers use MR fluid as damping medium. The properties of MR fluids were not conducive to the working of viscous dampers and thus these dampers could not be exploited commercially. In a recent work [4], we used a ferrofluid as a damping medium and showed that it is possible to control damping behavior of a ferrofluid based damper also using magnetic field. This paper reports results of a systematic study of the effect of damping efficiency of the damper as a function of magnetic field. It is believed that ferrofluid based dampers are better suited than MR dampers for regular use in buses and cars [4,5].

## II. FERROFLUID BASED DAMPERS

Conventional dampers used in cars or buses are piston-cylinder type viscous dampers where the motion of piston is damped by the viscous liquid present in the cylinder. The damping behavior of above damper can be controlled by changing the viscosity of damping medium. It seems, rheology of ferrofluids can be controlled by applying magnetic field [6] and thus these liquids are ideally suited for active dampers. Ferrofluids should not be confused with magnetorheological fluid (MR fluids), even though both of them are magnetic field-responsive fluids. Unlike ferrofluids which consist of nano-sized particles, MR fluids consist of micron sized particle suspended in a suitable oil. The magnetic particles tend to self-aggregate and thus MR fluids are not stable. Moreover, they exhibit hysteresis. On the other hand, ferrofluids are much stable and ferrofluid dampers have advantage of long life. The higher fluidity of ferrofluids as compared to MR fluids makes them better suited in damper applications.

## III. DAMPING EFFICIENCY OF A DAMPER

The piston of a piston-cylinder type of oscillator performs an oscillatory motion. The amplitude of oscillations is independent of time in a free oscillator. However, amplitude of oscillations decreases with time in a damped oscillator.

Damping efficiency of a viscous damper is measured in terms of a dimensionless parameter ζ (referred to as damping ratio), which describes how rapidly the oscillations decay from one bounce to the next. The damping ratio ζ is defined in terms of a parameter δ (logarithmic decrement) such that

$$\delta = \frac{1}{n} ln\left(\frac{x_1}{x_{n+1}}\right) \quad \text{and} \quad \zeta = \sqrt{\frac{\delta^2}{\delta^2 + 4\pi^2}} \qquad (1)$$

where $x_n$ is the amplitude of oscillation at the end of $n^{th}$ cycle. Depending on the value of ζ, the oscillatory motion is referred to as undamped (ζ = 0), underdamped (ζ < 1), critically damped (ζ = 1) or overdamped (ζ > 1). The measurement of damper efficiency or damper ratio ζ involves monitoring the time dependence of amplitude of oscillation.

## IV. EXPERIMENTAL DETAILS

Fig.1 is a schematic drawing of the piston-cylinder viscous damper used in present studies. Piston oscillates in a cylinder having diameter of 25.4 mm. The amplitude of oscillation of the piston was monitored as a function of time, and these data are used to calculate the damping efficiency or damping ratio of the damper. Measurements were made under varying experimental conditions. First, data were recorded for undamped damper when there was no viscous liquid in the cylinder. The other standard run was taken using water as the damping medium. The main data were taken using ferrofluid as damping medium and exposing it to varying magnetic fields. The damping medium was exposed to magnetic field using permanent neodymium magnets. There was a provision to move the magnets laterally and

This research is financially supported by *All India Council of Technical Education* (AICTE)—*New Delhi*, under *Research Proposal Scheme (RPS)*.

thereby change the magnetic field. The strength of magnetic field was measured using a hall probe.

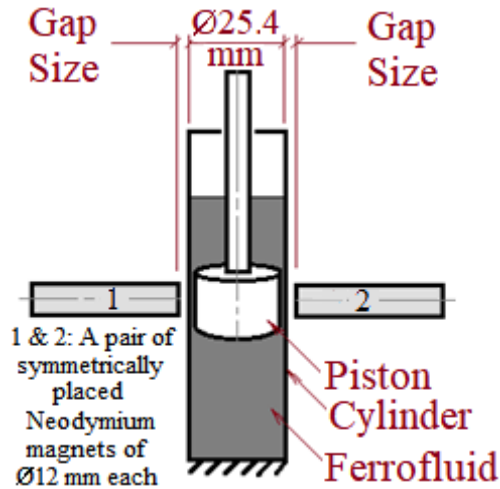

Fig. 1. Piston-cylinder damper using ferrofluid as damping medium. The fluid is exposed to magnetic field using permanent magnets.

The ferrofluid used for above studies was synthesized in our laboratory, the details of which are given in an earlier paper [4]. It consisted of a suspension of oleic acid coated $Fe_3O_4$ nanoparticles in kerosene. The said nanoparticles were synthesized using co-precipitation technique with $FeCl_3$ and $FeCl_2$ as starting materials.

The oscillation behavior of the piston was studied by connecting the piston to the vibrating system (Fig. 2), which consists of a beam pivoted at its one end and supported by a helical spring at the other end. Vibrations are sensed using an accelerometer sensor and its associated electronics [4]. The amplitude - time graphs are obtained from acceleration - time graphs using standard integrations. This is illustrated in Fig.3 where amplitude - time graph is obtained from acceleration - time graph for a typical data corresponding to magnetic field H = 3.2 kG. The effect of magnetic field H on the amplitude - time graphs have been studied for several values of H in range of 0.0 to 4.5 kG.

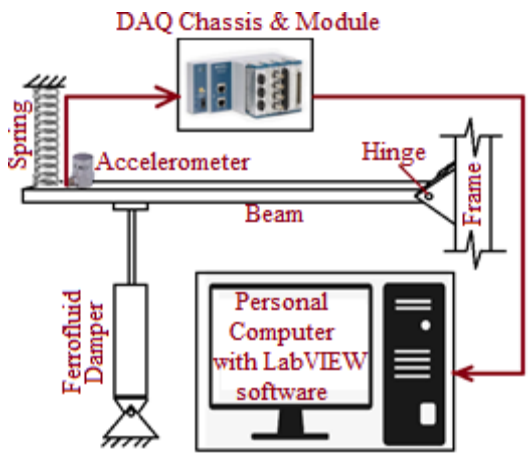

Fig. 2. Vibrating system used for measuring amplitude-time graphs for piston-cylinder damper. The acceleration is measured using uni-axial accelerometer and corresponding electronics.

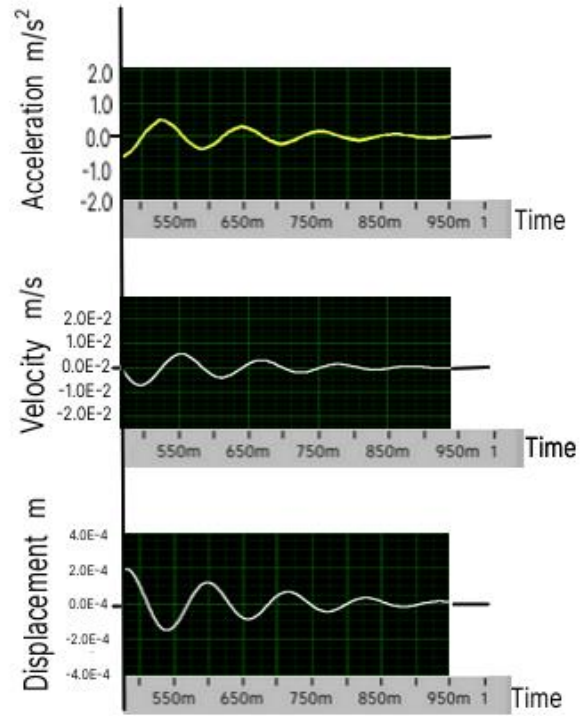

Fig.3 LabVIEW readings obtained for Displacement, Velocity and Acceleration versus Time, for a typical data corresponding to magnetic field H = 3.2 kG.

## V. RESULTS AND DISCUSSIONS

The magnetic field in region of ferrofluid depends on the position of magnets or the gap size between the magnet and the cylinder. Fig. 4 gives the variation of magnetic field as a function of gap size. This plot was obtained by measuring magnetic field using a hall probe. In actual experiments, measurements were made for different gap sizes and the above calibration curve was used to obtain the value of the magnetic field.

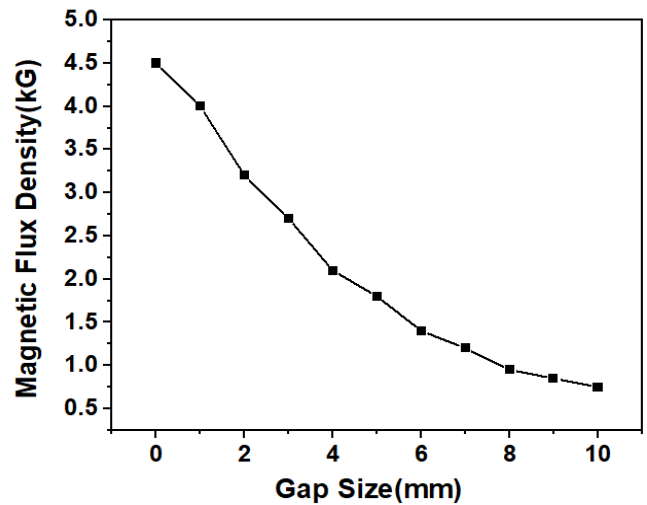

Fig.4 Calibration curve between gap size and magnetic field.

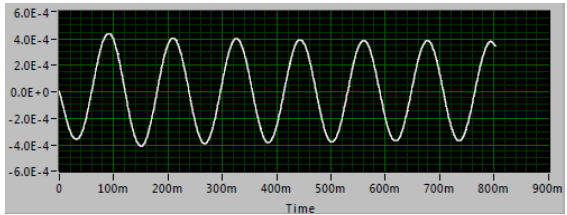
(i) Undamped (no damper connected)

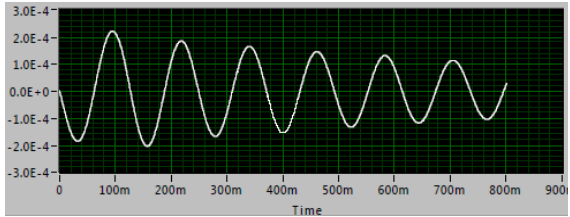
(ii) Damper with water as damping medium

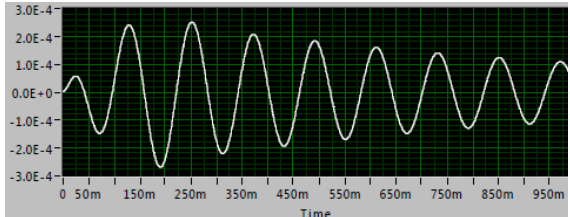
(iii) Ferrofluid based damper without magnetic field

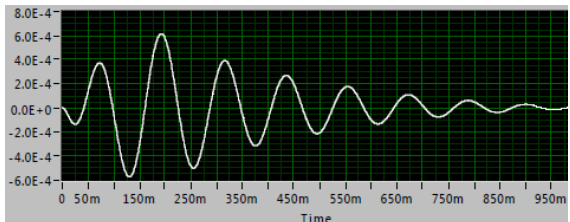
(iv) Ferrofluid based damper for H = 3.2 kG

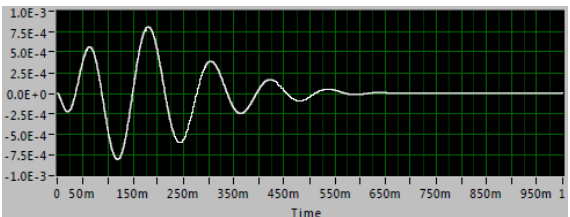
(v) Ferrofluid based damper for H = 4.5 kG

Fig. 5. Displacement versus Time graphs for the piston in viscous medium under different magnetic fields.

Fig. 5 shows the time dependence of amplitude of oscillations of the piston in above piston-cylinder damper corresponding to different magnetic fields (see Table 1). For the sake of completeness, amplitude- time graph of an undamped system is shown in Fig. 5 (i). It is noted that amplitude of vibrations does not change with time. Fig. 5 (ii) shows amplitude- time graph for a damper that uses water as damping medium. The amplitude of vibrations decreases slowly with time. Figures 5 (iii) to 5 (v) show amplitude- time graphs for a damper that uses ferrofluid as damping medium. It is noted that damping of amplitude of vibration for ferrofluid-based damper is much more than that for water-based damper. Figures 5 (iii), 5 (iv) and 5(v) correspond to measurements when damping medium was exposed to magnetic fields of 0.0 kG, 3.2 kG and 4.5 kG respectively. It is seen that the damping increases with increase in the magnetic field.

It may be noted that Fig. 5 shows typical data, but actual measurements have been made for a number of H values (Table 1). These data along were used to calculate damping ratio ζ using the formulae given in Section III. Results are shown in Table 1. Each measurement was repeated 5 times and average values of ζ are given in Table 1. Variation of ζ with magnetic field is shown in Fig.6. It is interesting that ζ varies linearly with H. Slope of curve suggests that rate of increase of damping efficiency is about 0.028 per kG. It seems the magnetic nanoparticles form chains in presence of magnetic field and that gives rise to increase in viscosity and the damping ratio.

TABLE I
EFFECT OF MAGNETIC FIELD ON DAMPING RATIO

| Sr. No. | Gap Size (mm) | Magnetic Flux Density (kG) | Average Damping Ratio corresponding to five readings |
|---|---|---|---|
| 1 | 100 mm | 0 | 0.020 |
| 2 | 10 | 0.75 | 0.043 |
| 3 | 9 | 0.85 | 0.046 |
| 4 | 8 | 0.95 | 0.043 |
| 5 | 7 | 1.2 | 0.050 |
| 6 | 6 | 1.4 | 0.054 |
| 7 | 5 | 1.8 | 0.056 |
| 8 | 4 | 2.1 | 0.067 |
| 9 | 3 | 2.7 | 0.076 |
| 10 | 2 | 3.2 | 0.091 |
| 11 | 1 | 4 | 0.099 |
| 12 | 0 | 4.5 | 0.148 |

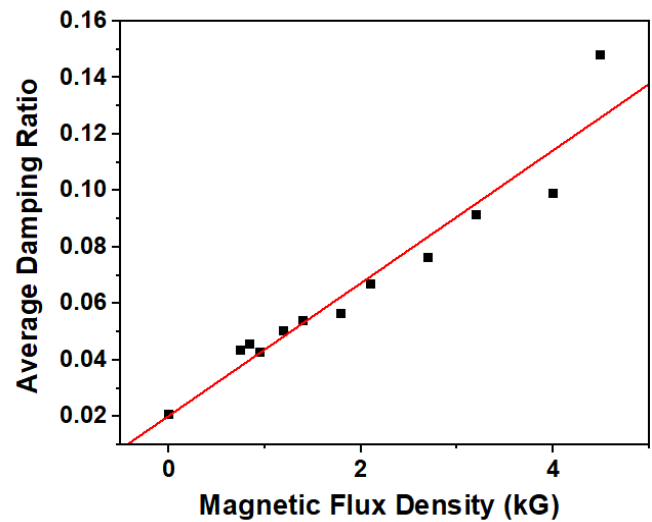

Fig. 6. Variation of damping ratio with magnetic field for ferrofluid based damper.

## VI. Conclusions

This paper deals with ferrofluid based damper, where one uses ferrofluid as the damping medium. In an earlier work, we had provided a proof of concept and showed that the damping behavior of the damper can be varied by applying magnetic field to the damping medium. The present paper reports a systematic study of the variation of damping ratio $\zeta$ with the magnetic field H. It is seen that $\zeta$ increases linearly with magnetic field up to H = 4.5 kG. The damping efficiency increases by about 600 % as the magnetic field is changed from H = 0.0 to H = 4.5 kG.

The fact that ferrofluids are more stable than MR fluids, it is expected that ferrofluid-dampers are better suited than MR Fluid-dampers for commercial exploitation.

## Acknowledgment

We thank Biswajit Panda for help in synthesizing ferrofluids and R I K Moorthy for useful discussions. Encouragement from Priam Pillai and Sandeep M Joshi is acknowledged.